\documentclass{IEEEtran}
\usepackage{cite}
\usepackage{amsmath,amssymb,amsfonts}
\usepackage{graphicx}
\usepackage{array}
\usepackage{xcolor}
\usepackage{multirow}
\usepackage{textcomp,nicefrac}
\def\BibTeX{{\rm B\kern-.05em{\sc i\kern-.025em b}\kern-.08em
T\kern-.1667em\lower.7ex\hbox{E}\kern-.125emX}}

\begin{document}

\title{Neutron Yield of Thermo Scientific P385 \\ D-T Neutron Generator \textit{vs.} Current and Voltage}
\author{Jihye~Jeon, Robert~J.~Goldston, and Erik~P.~Gilson
\thanks{This work was supported by the U.S. Department of Energy under contract number DE-AC02-09CH11466. The United States Government retains a non-exclusive, paid-up, irrevocable, world-wide license to publish or reproduce the published form of this manuscript, or allow others to do so, for United States Government purposes.} 
\thanks{Jihye Jeon is with Princeton University, Princeton, NJ 08542 USA (e-mail: jihyej@princeton.edu).}
\thanks{Robert J. Goldston is with Princeton University and Princeton Plasma Physics Laboratory, Princeton, NJ 08540 USA (e-mail: goldston@pppl.gov).}
\thanks{Erik P. Gilson is with Princeton Plasma Physics Laboratory, Princeton, NJ 08540 USA (e-mail: egilson@pppl.gov).}}

\maketitle

\begin{abstract}
The Thermo Scientific P385 Neutron Generator is a compact neutron source, producing 14 MeV neutrons through the deuterium-tritium (DT) fusion reaction. For practical use, it is important to measure and preferably understand the dependence of the neutron production rate on the accelerator current and voltage. In this study, we evaluated neutron production with a neutron spectrometer (BTI N-Probe), a He-3 detector surrounded by HDPE shells (Nested Neutron Spectrometer, NNS), and two ZnS fast neutron scintillators (EJ-410) for both P385 A3082 and A3083 sealed tubes. We also predicted the neutron yield using the TRIM code, which calculates the trajectory and the energy loss of deuterons and tritons within the target. Experimental and theoretical results showed a linear dependence on beam current and a $\sim$3.5 power law dependence on the operating voltage. A series of NNS measurements, neutron spectrum from N-Probe and MCNP calculation showed that the A3083 and A3082 tubes provide a maximum neutron yield of $\sim 8.2 \times 10^8$ n/s and $\sim 4.5 \times 10^8$ n/s, respectively. We showed that tritium decay was not a significant contributor to this difference.
\end{abstract}

\begin{IEEEkeywords}
accelerator-based neutron source, DT neutron generator, P385 neutron generator
\end{IEEEkeywords}

\section{Introduction}
\label{sec:introduction}
\IEEEPARstart{T}{he} The Thermo Scientific P385 Neutron Generator is a compact neutron source with a specified maximum neutron yield of $5 \times 10^{8}$ neutrons/s \cite{thermo}. The P385 creates 14 MeV neutrons by the deuteron-triton (DT) fusion reaction. In our laboratory, we are exploring its use in neutron radiography for arms control \cite{goldston}. The neutron yield of the generator is determined by the operating current and voltage of the accelerator. The accelerator head for P385 with DT neutron tube model A3082 is specified to have a neutron yield of $5 \times 10^8$ n/s at the maximum operating current of 70 $\mu$A and voltage of 130 kV. The A3083 tube is specified to provide $8 \times 10^8$ n/s at the maximum operating current of 90 $\mu$A and voltage of 140 kV. To better characterize the source for future applications, it is important to understand the dependence of neutron production rate on the accelerator current and voltage. 

In a study using another model of DT generator, GENIE 16, authors determined the voltage dependence, which was found to have a power law of 3.2, based on 11 different combinations of high voltage and current \cite{fondement}. The vendor of GENIE 16, SODERN, estimates the neutron flux to be proportional to the current and to follow a power law of 3.5--3.7 with respect to the voltage from 60 kV to 120 kV (GENIE16 User Manual). On the other hand, the Thermo Scientific P385 Operation Manual states that the neutron output depends on the high voltage with the power law relationship of 1.5 \cite{thermo2}. 

This paper provides the experimental measurements and theoretical analysis of the effect of the beam current and voltage of the P385 DT neutron generator. Furthermore, this paper confirms the absolute source rate of A3083 and A3082 tubes and determines whether tritium decay in the beam and target can explain the different neutron production rate from two neutron tubes at the same operating settings.

% -----------------------------------------------------------------------------------------------

\section{Neutron Yield Calculation}

For a model, we wrote a simple Python code to calculate the neutron yield as a function of incident beam energy. The fraction of incoming ions that undergoes fusion is given by
\begin{align*}
    f = n_t\sum_i{\Delta l_i ~\sigma_{\mathrm{fus},i}}
\end{align*}
where the $i$ denotes individual steps by the beam ions within the target material between its initial and final, low, energy, $n_t$ is the target density, $\Delta l_i = \sqrt{(\Delta x_i)^2 + (\Delta y_i)^2 + (\Delta z_i)^2}$ is the length of the $i$'th step, and $\sigma_{\mathrm{fus},i}$ is the fusion cross-section for the specified beam and target species, using the beam energy at the $i$'th step.

We calculated $\Delta l$ using the TRIM code (the Transport of Ions in Matter) \cite{srim}, which is a Monte Carlo program capable of calculating the 3D distribution of the ions in the target. With TRIM, we acquired the trajectory of 10,000 deuteron and triton ions, including both longitudinal and transverse straggling, at each energy of our interest, providing steps with the energy interval of 100 eV. Then, we calculated $\Delta l$ between steps. We used the ENDF/B-VIII.0 library to interpolate the fusion cross section, $\sigma_{\mathrm{fus},i}$, at the corresponding energy of the beam species, assuming the target species was at rest. We calculated the fusion rate of the individual 10,000 ions with the same initial energy using the equation and then acquired the average value. 

Based on communication with Thermo Fisher Scientific\cite{simpson}, we assumed that the target is $\sim$1.5 $\mu$m thick titanium metal loaded with $\sim$1.85 hydrogenic atoms per titanium atom so that we would have $n_T = n_D = 0.925 n_{Ti}$ where $n_T$, $n_D$ and $n_{Ti}$ are the number density of tritium, deuterium and titanium atoms in the target.  

According to the Thermo Fisher Scientific\cite{simpson}, the beam is estimated to be 90\% diatomic ($DT^+$, $TT^+$, $DD^+$) and 10\% monatomic ($D^+$, $T^+$). Any triatomic component, e.g., $DDT^+$, is negligible. The atomic (or monatomic) and molecular beam give us different neutron production yields. Since even with the same acceleration energy, for example, a deuteron atom in $DD^+$ molecular beam will have half energy of a deuteron in monatomic $D^+$ beam.

In a monatomic beam, we expect the $D^+$ or $T^+$ to have energy $eV_{\mathrm{acc}}$ where $V_{\mathrm{acc}}$ is the full accelerator voltage. Thus the maximum energy that either species can have in our experiments is $eV_{\mathrm{acc,max}}$, where $eV_{\mathrm{acc,max}}$ is the maximum applied accelerator voltage. In a diatomic beam both nuclei are moving at the same velocity. Thus in a diatomic beam, the nuclei share the energy proportional to their atomic mass, as shown in Table~\ref{tab:mix&match}.

We have a beam electrical current of $I_{\mathrm{acc}}$, but we need  ``currents'' of nuclei. We assume that $D$ and $T$ are evenly mixed everywhere and then we have that the full energy atomic $D$ and $T$ beams are equal so $D^+$ and $T^+$ ions take half of 10\% of total current. When it comes to diatomic beams, simple combinatorial arguments indicate that 1/2 of 90\% of total current should be delivered in the form of $DT^+$ and 1/4 in the each of the forms of $DD^+$ and $TT^+$. The energy and current of each atom in the beam is summarized in Table~\ref{tab:mix&match}.

% ********************

\begin{table}[!htb]
\caption{Energy and current of beam particles with the accelerator voltage $V_{\mathrm{acc}}$ and current $I_{\mathrm{acc}}$.}
\label{tab:mix&match}
{\begin{tabular}{|m{0.5cm} m{3.3cm} m{3.7cm}|}
\hline
Ion & Energy & Current\\
\hline \hline
\multicolumn{3}{|l|}{10\% atomic beam} \\
$D^+$ & $eV_{\mathrm{acc}}$ & $0.05I_{\mathrm{acc}}$ \\
$T^+$ & $eV_{\mathrm{acc}}$ & $0.05I_{\mathrm{acc}}$ \\
\hline \hline
\multicolumn{3}{|l|}{90\% molecular beam} \\
$DD^+$ & $0.5eV_{\mathrm{acc}}$ ($D$), $0.5eV_{\mathrm{acc}}$ ($D$) & $0.225I_{\mathrm{acc}}$ ($D$), $0.225I_{\mathrm{acc}}$ ($D$) \\
$TT^+$ & $0.5eV_{\mathrm{acc}}$ ($T$), $0.5eV_{\mathrm{acc}}$ ($T)$ & $0.225I_{\mathrm{acc}}$ ($T$), $0.225I_{\mathrm{acc}}$ ($T$) \\
$DT^+$ & $0.4eV_{\mathrm{acc}}$ ($D$), $0.6eV_{\mathrm{acc}}$ ($T$) & $0.45I_{\mathrm{acc}}$ ($D$), $0.45I_{\mathrm{acc}}$ ($T$) \\
\hline
\end{tabular}}
\end{table}
% ********************

In the calculation, this breakdown on the current of nuclei was taken into account as a weighting factor for the linear combination of each atomic and molecular beam.

% ------------------------------------------------------------------------------
\section{Experiments}

\begin{figure}[!htb]
    \centering
    \includegraphics[height=2.2in]{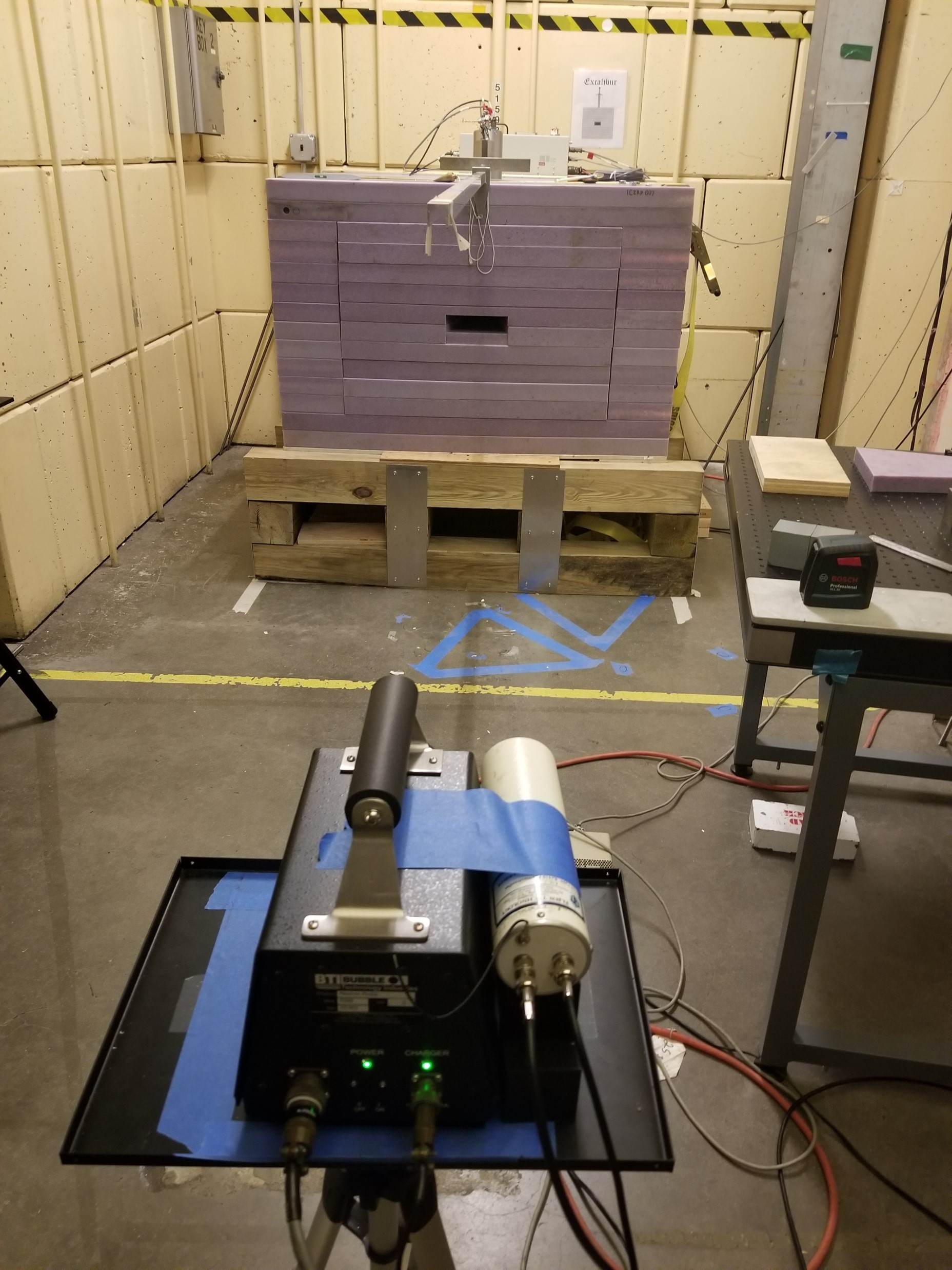}
    \includegraphics[height=2.2in]{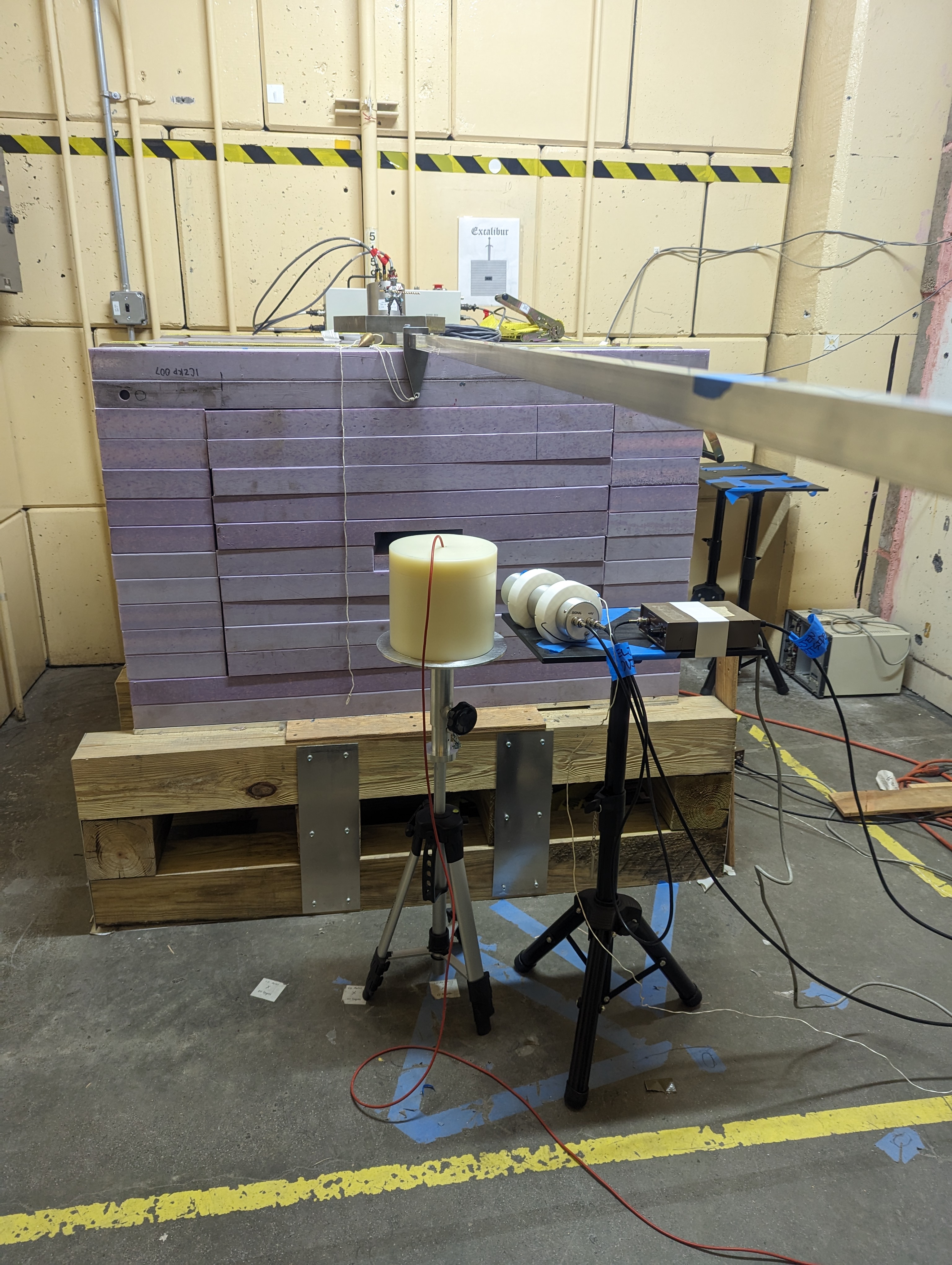}
    \caption{Experimental configuration. P385 neutron source is surrounded by a carbon steel and borated polyethylene collimator \cite{hepler}. A neutron exit aperture that has a tapered shape in the horizontal plane is created in the steel cylinder and polyethylene block with an opening angle of 17.5 degrees and a height of 2$^{\prime\prime}$. The left figure shows the BTI N-Probe and a EJ-410 detector at 3 m distance. The right figure shows the NNS and an EJ-410 detector placed at 1 m distance from the source. In both configurations, another EJ-410 detector is on the floor under the P385 and the collimator, covered by a layer of 2$^{\prime\prime}$-thick lead bricks, which is not shown in the picture. Between the P385 and the EJ-410 on the floor, there is a cylindrical cavity in the collimator and moderator, directed towards this detector.}
    \label{fig:exp_setting}
\end{figure}

We conducted a series of experiments to measure the dependence of neutron production on the intensity of neutron beam. The absolutely calibrated neutron detectors are a BTI N-Probe (N-Probe, Bubble Technology Industries, Inc.) and a Nested Neutron Spectrometer (NNS, Detec, Inc.) as shown in Figure~\ref{fig:exp_setting}.

The BTI N-Probe is a commercial neutron spectrometer which measures neutrons from thermal energies up to 20 MeV using a $^3He$ proportional counter for energies below 800 keV and a liquid scintillator for energies above 800 keV \cite{ing}. The N-Probe, coupled with BTI MICROSPEC analyzer, provides a neutron spectrum by spectral unfolding \cite{koslowsky}. The BTI N-Probe was used to count neutrons for current dependence measurements and provide a neutron spectrum for absolute source rate analysis. The N-Probe was placed 3~m away from the target of the P385 where 14 MeV neutrons are generated.

The second detector, NNS, consists of seven high density polyethylene (HDPE) cylindrical shells and a $^3He$ proportional counter. The HDPE shells are assembled in a nested fashion so that eight independent measurements, from the bare $^3He$ detector to all the seven shells, provides the neutron energy spectrum from 1.56 meV to 20 MeV by unfolding \cite{detec}. Theoretically, with those eight measurements, one can get a neutron spectrum unfolded by the NNS response functions. However, the projection of the collimated beam at 1~m is $\sim$8.3 cm high and this will not cover the full height of the 3rd smallest shell, which is 8.636 cm. Therefore, we deemed that this collimated source setting is not appropriate to use the built-in response function of the NNS. For voltage dependence measurements, the NNS was left at 1~m distance from the source with six HDPE shells for all the time with different operating voltages since this setup gives us the most neutron counts among the eight measurements. For absolute source rate experiments, we measured neutrons with the NNS with all 7 shells and then subsequently removed one shell around the detector and counted neutrons for each configuration, as in the conventional NNS measurements, except not using the unfolding process, but comparing to MCNP calculations. 

For both the N-Probe and the NNS, a 2$^{\prime\prime}$ ZnS fast neutron detector (EJ-410, Eljen Technology, Inc.) was always placed next to N-Probe or NNS.

\section{Results}

\subsection{Current and voltage dependence}

\begin{figure}[!htb]
    \centering
    \includegraphics[width=2.7in]{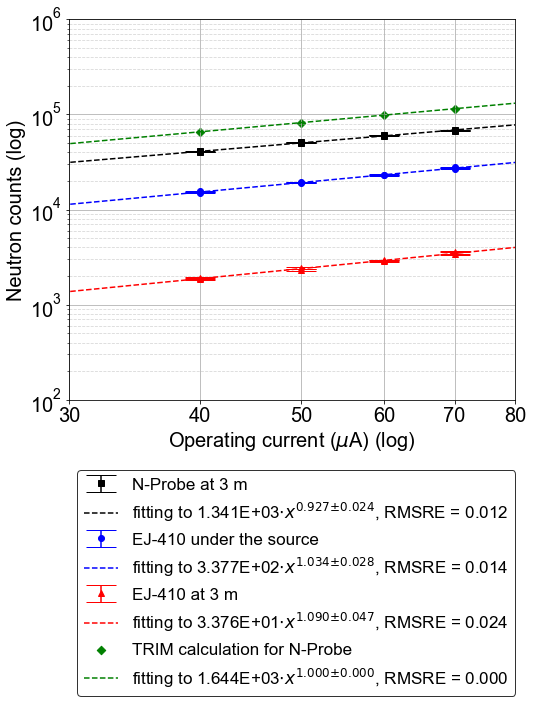}
    \caption{Experimental results on current scanning of P385 A3082. The A3082 tube was operated at its maximum voltage of 130 kV. Instead of using the neutron flux from the unfolded spectrum of the N-Probe, we used the pulse height spectrum which has less scatter. The energy range of 4--10 MeV is chosen where it shows a counting plateau in the integral pulse height spectra, providing maximum stability over long periods of time \cite{knoll}. The root mean squared relative error (RMSRE) for each fitting is noted.}
    \label{fig:current}
\end{figure}

Figure~\ref{fig:current} shows the current dependence of the A3082 neutron emission rate at the operating voltage of 130 kV. When fitted with a power law, the neutron counts from the N-Probe and two EJ-410 fast neutron detectors had scaling exponents close to unity with the P385 tube current over the working range 40--70~$\mu$A. The EJ-410 detector at 3 m showed the most deviation from linearity with a power of 1.09 ($\pm 0.05$) and the one under the source presented the most linearity with a power of 1.03 ($\pm 0.03$).

\begin{figure}[!htb]
    \centering
    \includegraphics[width=2.7in]{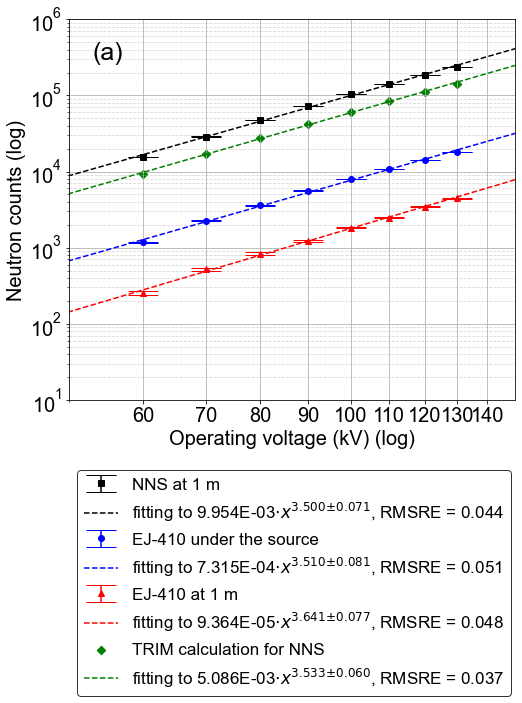}
    
    \vspace{1cm}
    
    \includegraphics[width=2.7in]{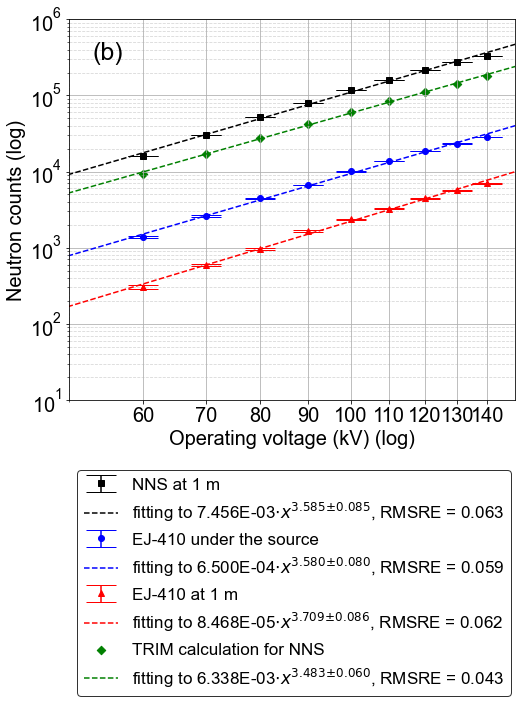}
    \caption{Theoretical and experimental results on voltage scanning of (a) A3082 and (b) A3083. The plot on the left shows the voltage dependence of the tube A3082 while the one on the right shows that of A3083 when both tubes were operated at the current of 70~$\mu$A. The root mean squared relative error (RMSRE) for each fitting is noted.}
    \label{fig:voltage}
\end{figure}

With regard to voltage dependence at a current of 70~$\mu$A, we found that the neutron flux scales like the 3.50--3.64 power of the tube voltage over the working range 60--130 kV for the tube A3082 and 3.58--3.71 power over 60--140 kV for the A3083 as shown in Figure~\ref{fig:voltage}. These power law exponents match with the GENIE 16 user manual (power law exponent of 3.5--3.7) and the unpublished study (power law exponent of 3.61 -- 3.74) \cite{chichester}. This study used Thermo Scientific MP 320 Neutron Generator and measured the neutron yield under the current of 40 -- 60 $\mu$A and the voltage of 50 -- 90 kV. We could fit the curve on the voltage by the power law of 3.61 -- 3.74. The TRIM calculation was in good agreement with the measurement, showing the power of 3.53 and 3.48 for the A3082 and A3083, respectively, where the small difference is due to the range of voltages tested. The EJ-410 scintillator under the source, covered with a lead block, and the NNS showed close power scaling. Their power exponents only differ by 0.28\% for A3082 and 0.14\% for A3083.

% ------------------------------------------------------------------------------

\subsection{Absolute source rate}

\begin{figure}[!htb]
    \centering
    \includegraphics[width=2.7in]{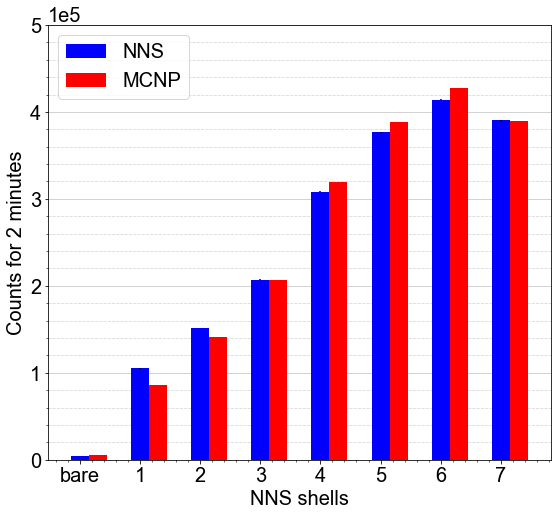}
    \caption{Experiment (NNS) and theoretical (MCNP) He-3 neutron counts. Neutrons were measured for two minutes at a 1 m distance from the source, A3083, at the current of 90 $\mu$A and voltage of 140 kV. MCNP results were scaled by the estimated source rate, $\sim 8.23 \times 10^8~\mathrm{n/s}$. 
    }
    \label{fig:nns_abs}
\end{figure}

We estimated the absolute source rate at the maximum operating setting of the A3082 and A3083. First, the full set of the NNS measurements of the A3083 tube was performed at its maximum setting, current of 90 $\mu$A and voltage of 140 kV. The NNS has been calibrated with an AmBe standard source of the Ionizing Radiation Standards Laboratory at the National research Council of Canada (IRS-NRC) \cite{detec}. A Monte Carlo calculation by MCNP6 \cite{werner} was conducted for each shell configuration, to determine the corresponding absolute emission. Weighting eight measurements by their neutron counts, the maximum source rate of the A3083 was estimated to be $\sim 8.2 \times 10^8~\mathrm{n/s}$. Figure~\ref{fig:nns_abs} compares the neutron counts from the eight measurements from experiment and calculation after taking into account the estimated A3083 source rate.

The maximum neutron emission from the A3082 has been measured with the NNS surrounded by six shells, which have the highest neutron counts and the lowest statistical uncertainties among eight measurements. Based on the MCNP6 calculation, at the current of 70 $\mu$A and voltage of 130 kV, the A3082 tube provided $\sim 4.5 \times 10^8~\mathrm{n/s}$.

Another neutron detector, the N-Probe, has been used to cross check the results from the A3082. The N-Probe was located at a 3 m distance from the source with its maximum neutron yield. The flux above 6 MeV in the spectrum is $\sim 396~\mathrm{n/cm^2/s}$. Considering the detector distance from the source, we can estimate the source rate of $\sim 4.5 \times 10^8~\mathrm{n/s}$. The MCNP calculation showed that the source neutron should have a rate of $\sim 4.6\times 10^8~\mathrm{n/s}$ in order to match the neutron flux above 6 MeV from the experiment.

TRIM calculation gave $1.2 \times 10^9~\mathrm{n/s}$ for A3082 maximum neutron yield and $1.5 \times 10^9~\mathrm{n/s}$ for A3083. This is a factor of 2.6 (A3082) and 1.8 (A3083) greater than the yield estimated from the experiments. This suggests the presence of impurities in the Ti material, displacing $D$ and $T$ nuclei.

% ------------------------------------------------------------------------------

\subsection{Effect of tritium decay}

Since we have two accelerator tubes, A3082 ($\sim$8.44 years from the manufacture date) and A3083 ($\sim$1.44 years), it provides an opportunity to understand the temporal degradation of the neutron generator. If two tubes share the same configuration and initial state, then the difference in neutron counts at the same operating setting can give us a hint of how the source rate gradually changes over time, for example, due to tritium decay. Therefore, we focused on calculating the tritum decay in this study.

Tritium has a half-life of 12.3 years and decays to $^3He$. We assumed that $^3He$ generated from tritium decay is not trapped in the titanium target and that the number of hydrogenic atoms per titanium atom is conserved by the beam. We supposed that the target is in equilibrium with the surrounding gas from which the beam is generated. Therefore, we would have the same ratio of deuterons to tritons in the beam and target. 

\begin{figure}[!htb]
    \centering
    \includegraphics[width=2.7in]{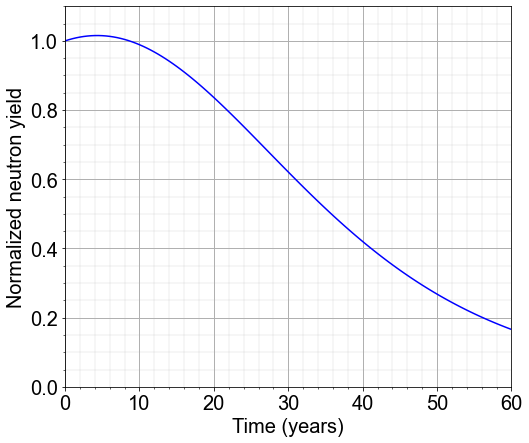}
    \caption{Neutron yield at 70 $\mu$A and 130 kV over the time. The yield is normalized by the yield at the date of manufacture.}
    \label{fig:tritium}
\end{figure}

The fraction in tritons $t$ years after the date of manufacture, $f_T(t)$, can then be expressed as,
\begin{align*}
    f_T(t) = \dfrac{f_T(t_0)\cdot e^{-\lambda t}}{f_T(t_0)\cdot e^{-\lambda t}+f_D(t_0)}
    = \dfrac{e^{-\lambda t}}{e^{-\lambda t}+1}
\end{align*}

where $\lambda = \ln(2)/12.3$. We assumed there are equal number of deuterons and tritons in the gas when the generator is manufactured $(f_T(t_0) = f_D(t_0) = 0.5)$. The deuteron fraction is simply $f_D(t) = 1-f_T(t)$. We can have a generalized model by incorporating this temporal change to the current in Table~\ref{tab:mix&match} as shown in Table~\ref{tab:mix&match2}. The calculation result of the fusion reaction rate at the voltage of 130 kV is shown in Figure~\ref{fig:tritium}.

\begin{table}[!htb]
\caption{Energy and current of beam particles with the accelerator voltage $V_{\mathrm{acc}}$ and current $I_{\mathrm{acc}}$ after $t$ years from the manufacture date.}
\label{tab:mix&match2}
{\begin{tabular}{|m{1.5cm} m{3cm} m{3cm}|}
\hline
Ion & Energy & Current\\
\hline \hline
\multicolumn{3}{|l|}{10\% atomic beam} \\
$D^+$ & $eV_{\mathrm{acc}}$ & $0.1 f_D(t) I_{\mathrm{acc}}$ \\
$T^+$ & $eV_{\mathrm{acc}}$ & $0.1 f_T(t) I_{\mathrm{acc}}$ \\
\hline \hline
\multicolumn{3}{|l|}{90\% molecular beam} \\
\multirow{2}{1cm}{$DD^+$} & $0.5eV_{\mathrm{acc}}$ ($D$) & $0.45 f_D(t) I_{\mathrm{acc}}$ ($D$) \\
& $0.5eV_{\mathrm{acc}}$ ($D$) & $0.45 f_D(t) I_{\mathrm{acc}}$ ($D$) \\
\hline
\multirow{2}{1cm}{$TT^+$} & $0.5eV_{\mathrm{acc}}$ ($T$) & $0.45 f_T(t) I_{\mathrm{acc}}$ ($T$) \\
& $0.5eV_{\mathrm{acc}}$ ($T$) & $0.45 f_T(t) I_{\mathrm{acc}}$ ($T$) \\
\hline
\multirow{2}{1cm}{$DT^+$} & $0.4eV_{\mathrm{acc}}$ ($D$) & $0.9 f_D(t) I_{\mathrm{acc}}$ ($D$) \\
& $0.6eV_{\mathrm{acc}}$ ($T$) & $0.9 f_T(t) I_{\mathrm{acc}}$ ($T$) \\
\hline
\end{tabular}}
\end{table}
% ********************

Based on the calculation, we would get the neutron yield higher than its initial yield until 8.5 years with a maximum around 4.5 years. This is not surprising since the beam would have the same total number of ions if we keep running the same current and due to tritium decay we would have a higher ratio of deuterons in the beam. Since the main contributors to the neutron yield are $D^+$ and $DD^+$ going under D-T reaction, a greater number of deuterons in the beam would at first increase the D-T reaction rate. Over the time, lower portion of tritons in the target would decrease the reaction rate. This gives a weaker response to T decay than might have been naively expected, by just considering the $D$ and $T$ densities in the beam and target.

At the time on the experiment, the A3083 was $\sim$1.44 years after the date of manufacture and the A3082 $\sim$8.44 years. Based on the curve in Figure~\ref{fig:tritium}, the fusion rate of the A3083 increases by 0.85\% and the A3082 by 0.13\% from their initial state. The ratio between these two tubes is 1.0072. The experiment showed that the ratio of neutron counts at the voltage of 130 kV between the A3083 and A3082 was 1.30 by the EJ-410 under the source, 1.27 by the EJ-410 at 1 m distance, and 1.17 by the NNS. This discrepancy possibly comes from different design between the two tubes, $^3$He pile-up, or different levels of impurities in the target material. Above all, target sputtering would affect the neutron yield the most and eventually the lifetime of the tube. \cite{simpson2} In addition, this analytical model could not explain the result that the ratio of neutron yield from the A3083 to the A3082 increased over the voltage --- from 3\% at 60 kV to 17\% at 130 kV when measured with NNS. 

% ------------------------------------------------------------------------------

\section{Conclusions}

In this study, we found the operating current and voltage dependence of P385 A3082 and A3083 neutron source. We measured neutrons with the BTI N-Probe, the NNS and two ZnS fast neutron scintillators. Also, we estimated the neutron yield with TRIM code, which calculates the trajectory and the energy loss of deuteron and triton ions within the target. Experimental and theoretical results showed the linear dependence to the operating current and $\sim$3.5 power dependence to the voltage. N-Probe, NNS measurements and MCNP calculation showed that the A3083 and A3082 tubes can give us a maximum neutron yield of $\sim 8.2 \times 10^8$ n/s and $\sim 4.5 \times 10^8$ n/s, respectively. We showed that these tubes should be less affected by tritium decay than would be naively expected, and that the lower performance of the older A3082 tube could not be explained by tritium decay.

% ------------------------------------------------------------------------------

\section*{Acknowledgment}

We would like to thank Mr. James Simpson from Thermo Fisher Scientific for providing information about Thermo Scientific P385 Neutron Generator and comments on this paper, Dr. David Chichester from Idaho National Laboratory for sharing his unpublished work relevant to our study and Dr. Valentin Fondement from University of Michigan for sharing his expertise on a similar D-T neutron generator, GENIE 16.

% ------------------------------------------------------------------------------


\begin{thebibliography}{00}

\bibitem{thermo} ``Thermo Scientific P 385 Product Specifications,'' Thermo Fisher Scientific, Colorado Springs, CO, USA.

\bibitem{goldston} R. Goldston, F. d'Errico, A. di Fulvio, A. Glaser, S. Philippe, and M. Walker, ``Zero Knowledge Warhead Verification: System Requirements and Detector Technologies,'' presented at the \textit{INMM Annual Meeting,} Atlanta GA, USA, July 20--24, 2014.

\bibitem{fondement} V. Fondement, B. P\'erot, T. Marchais, J. Loridon, H. Toubon, Y. Bensedik, and J. Collot, ``Electronics and 3He Counter Acquisition Tests During the Pulses of a D-T Neutron Generator,'' in \textit{Proc. IEEE NSS/MIC,} Milano, Italy, 2022,~pp.~1--6, 10.1109/NSS/MIC44845.2022.10399200.

\bibitem{thermo2}  ``Thermo Scientific P385 Neutron Generator Operation Manual,'' Thermo Fisher Scientific, Colorado Springs, CO, USA, Manual P/N 120006-A062907, 2010.

\bibitem{simpson} J. Simpson, private communication, December 2023.

\bibitem{srim} J.F. Ziegler, M.D. Ziegler, M.D., and J.P. Biersak, ``SRIM – The stopping and range of ions in matter (2010),'' \emph{NIM B}, vol. 268(11--12), pp. 1818--1823, June 2010, 10.1016/j.nimb.2010.02.091.

\bibitem{hepler} M. Hepler, ``Zero-knowledge Isotopic Discrimination for Nuclear Warhead Verification,'' Ph.D. disseration, Dept. Mech. Aero. Eng., Princeton Univ., Princeton, NJ, USA, 2020, pp. 67--70.

\bibitem{ing} H. Ing, S. Djeffal, T. Clifford, R. Machrafi, and R., Noulty, ``Portable spectroscopic neutron probe,'' \emph{Radiat. Prot. Dosimetry}, vol. 126(1--4), pp. 238--243, August 2007, 10.1093/rpd/ncm049.

\bibitem{koslowsky} M. Koslowsky, ``Spectral Unfolding: A Mathematical Perspective, Modern Neutron Detection,'' in \textit{Proc. a Technical Meeting by IAEA,} Vienna, Austria, IAEA-TECDOC-1935, 2020, pp. 97--115.

\bibitem{detec} ``Nested Neutron Spectrometer Operations, Data Acquisition and Data Analysis User Manual,'' ver. 2.1, Detec, Gatineau, QC, Canada.

\bibitem{knoll} G.F. Knoll, \textit{Radiation of Detection and Measurement}, 4th ed., John Wiley \& Sons, Inc., Hoboken, NJ, 2020, pp. 113--115.

\bibitem{chichester} D.L. Chichester, E.H. Seabury, and A.J. Caffrey, ``Measurement of the Neutron Yield of DD and DT Neutron Generators,'' presented at \textit{2008 Symposium on Radiation Measurements and Applications}, Berkeley, CA, USA, June 2--5, 2008. 

\bibitem{werner} C.J. Werner, J.S. Bull, C.J. Solomon, F.B. Brown, G.W. McKinney, M.E.
Rising, D.A. Dixon, R.L. Martz, H.G. Hughes, L.J. Cox, A. J. Zukaitis, J.C.
Armstrong, R. A. Forster, and L. Casswell, ``MCNP6.2 Release Note'', LANL, Los Alamos, NM, USA< Tech. Rep. LA-UR-18-20808, 2018.

\bibitem{simpson2} J. Simpson, private communication, May 2024.

\end{thebibliography}
\end{document}